\documentclass[apj]{emulateapj}

\usepackage{times}
\usepackage{graphicx}
\usepackage{amssymb}
\usepackage{multirow}
\usepackage{color}
\usepackage{wrapfig}
\usepackage{float}
\usepackage{apjfonts}
\usepackage[flushleft]{threeparttable}

\usepackage{lineno}
\usepackage{amsmath}
\usepackage{url}

\newif\ifAMStwofonts
\AMStwofontstrue

\makeatletter

\newcommand{\Rmnum}[1]{\expandafter\@slowromancap\romannumeral #1@}
\makeatother

\shorttitle{Annular Gap Model for MSPs}

\shortauthors{Du et al. 2012}

\slugcomment{Accepted for publication in The Astrophysical Journal}

\begin{document}


\title{Radio and Gamma-ray Pulsed Emission from Millisecond Pulsars}

 \author{Y.~ J.~ Du\altaffilmark{1,2},~
 G.~ J.~ Qiao\altaffilmark{3} 
 ~and~
 D.~ Chen\altaffilmark{1}}

 \altaffiltext{1}{National Space Science Center, Chinese Academy of
   Sciences, NO.1 Nanertiao, Zhongguancun, Haidian district, Beijing
   100190, China; duyj@nssc.ac.cn}

 \altaffiltext{2}{National Astronomical Observatories, Chinese Academy
   of Sciences, jia 20 Datun Road, Beijing 100012, China}

 \altaffiltext{3}{School of Physics, Peking University, Beijing
   100871, China}




\begin{abstract}
Pulsed $\gamma$-ray emission from millisecond pulsars (MSPs)
has been detected by the sensitive {\it Fermi}, which sheds
light on studies of the emission region and mechanism. In
particular, the specific patterns of radio and $\gamma$-ray emission
from PSR J0101-6422 challenge the popular pulsar models,
e.g. outer gap and two-pole caustic (TPC) models.
Using the three dimension (3D) annular gap model, we have jointly
simulated radio and $\gamma$-ray light curves for three representative
MSPs (PSR J0034-0534, PSR J0101-6422 and PSR J0437-4715) with distinct
radio phase lags and present the best simulated results for these
MSPs, particularly for PSR J0101-6422 with complex radio and
$\gamma$-ray pulse profiles and for PSR J0437-4715 with a radio
interpulse. It is found that both the $\gamma$-ray and radio emission
originate from the annular gap region located in only one
magnetic pole, and the radio emission region is not primarily lower
than the $\gamma$-ray region in most cases.  In addition, the annular
gap model with a small magnetic inclination angle instead of
``orthogonal rotator'' can account for MSPs' radio interpulse with a
large phase separation from the main pulse.
The annular gap model is a self-consistent model not only for young
pulsars but also MSPs, and multi-wavelength light curves can be
fundamentally explained by this model.
\end{abstract}

\keywords{pulsars: general --- gamma rays: stars --- pulsars: individual
  (J0034-0534, J0101-6422, J0437-4715) --- radiation mechanisms:
  non-thermal --- acceleration of particles}

\section{Introduction}

MSPs are a population of old neutron stars with short spin period $P_0
\lesssim 20$~ms and small characteristic magnetic field $B_0 < 10^9$~G
(actually small period derivative $\dot{P} \lesssim 10^{-17}\,{\rm
  s~s^{-1}}$). They are believed to be formed from the recycling
(accretion spin-up ) process in binaries \citep{msp}. Furthermore,
there is another possible formation channel for MSPs (even for
sub-millisecond pulsars), that is accretion-induced collapse of white
dwarfs in a binary \citep{nomoto91,smsp}. Because of the high
stability of their rotations, MSPs have great potential for
application: autonomous deep-space navigation, pulsar-based time
scale, low-frequency gravitational wave detection and so on.

When the first MSP, PSR B1937+21, had been discovered in 1982
\citep{backer82}, \cite{usov83} soon predicted that this pulsar could
emit $\gamma$-rays via synchrotron radiation on the order of 100
GeV. \cite{bhat91} subsequently calculated the $\gamma$-ray luminosity
of MSPs, estimated their contribution to the diffuse $\gamma$-ray
background of the Milky Way, and finally discussed detectability of
MSPs as point $\gamma$-ray sources.
Using the data of {\it Energetic Gamma Ray Experiment},
\cite{kuiper00} showed circumstantial evidence for the likely
detection of pulsed $\gamma$-ray emission from a MSP, PSR J0218+4232,
which was regarded as a $\gamma$-ray pulsar candidate for a long
time.
MSPs' $\gamma$-ray emission was not observationally confirmed
  until {\it Fermi} with a sensitive Large Area Telescope (LAT)
launched in June of 2008. Using eight-month data of {\it Fermi} LAT,
eight $\gamma$-ray MSPs have already been detected
\citep{msp-sci}. This number has grown to more than 40 up to date
\citep{guillemot}, and still increases.

From observations, MSPs are analogous to young pulsars, which have
multi-wavelength pulsed emission from radio to $\gamma$-ray band. Do
MSPs and young pulsars share a simple model that contains similar
emission region and acceleration mechanism to self-consistently
explain their multi-wavelength emission?
More and more multi-wavelength data with high precision give us
opportunities to obtain remarkable insights of the magnetospheric
physics. The multi-wavelength study is a key method to discriminate
the various pulsar non-thermal emission models for both MSPs and young
pulsars.

Initially aiming to explain the high-energy pulsed emission from young
pulsars, four traditional magnetospheric gap models have been
suggested to study pulsed high energy emission of pulsars: the polar
cap model \citep{1994ApJ...429..325D}, the outer gap model
\citep{1986ApJ...300..500C, 1995ApJ...438..314R, 1997ApJ...487..370Z},
TPC/slot gap model \citep{2003ApJ...598.1201D, harding08}, and the
annular gap model \citep{qiao04, qiao07, AG10}. To distinguish these
pulsar models, the most important issues are the inducements of
acceleration electric field region and related emission mechanisms to
emanate high-energy photons \citep{vela, crab}. One of the key
discrepancies of the mentioned emission models is one-pole or two-pole
emission pattern which depends on two corresponding geometry
parameters: magnetic inclination angle $\alpha$ and viewing angle
$\zeta$.

\citet{bulik00} adopted the polar-cap model to calculate the
$\gamma$-ray emission from MSPs. They pointed out that curvature
radiation of primary particles contributed to the MeV-to-GeV band,
while the synchrotron radiation arising from pairs dominated only
below 1 MeV. \cite{harding05} developed the pair-starved polar cap
model and obtain similar spectral conclusion for high-energy emission
from MSPs as above. In this model the accelerating field is not
screened and the entire open volume is available for particle
acceleration and emission of gamma rays.
\cite{zhang03} used the outer gap model with multi-pole magnetic field
to model the X-ray and $\gamma$-ray spectra for four MSPs, and the
predicted results basically agree with the observations
\citep{harding05}.

Along with radio observations supplying us with excellent timing
solutions for {\it Fermi} MSPs, the derived $\gamma$-ray and radio
light curves with high signal-to-noise allow us to do joint simulation
which can justify the pulsar emission models.
Recently, \citet{venter09} simulated both radio and $\gamma$-ray light
curves for MSPs in the pair-starved polar cap, TPC and outer gap
models, and they found that most of their simulated light curves are
well explained by the TPC and outer gap models. They especially
simulated light curves for a minor group of MSPs with phase-aligned
radio and $\gamma$-ray pulse profiles \citep{venter12}.
In addition, \cite{johnson12} also used the geometric slot gap, outer
gap model or pair-starved polar cap model to fit $\gamma$-ray
and radio light curves for three MSPs.

A MSP PSR J0101-6422 with complex radio and $\gamma$-ray light curves,
challenge the popular TPC and outer gap models \citep{kerr12}. It
  is found that neither of the two models can faithfully reproduce the
  observed light curves and phase lags. For such a complex radio
  profile the simple beam model they used may have been insufficient.

In this paper, we use a 3D annular gap model to study both radio
and $\gamma$-ray light curves for three MSPs which stand for the
relevant types of MSPs with different radio lags. In \S\,2, the
annular gap and core gap will be simply introduced and the
acceleration electric field in the annular gap is calculated. In
\S\,3, we jointly simulate radio and $\gamma$-ray band light
curves for PSR J0034-0534, J0101-6422 and J0437-4715. The radio phase
lags are identified and the reasons for them explored. The
conclusions and discussions are shown in \S\,4.

\section{The Annular Gap and Core Gap for MSPs}

\subsection{The Definition of Annular Gap and Core Gap}

In a pulsar magnetosphere \citep{GJ69,RS75}, the critical field
lines\footnote{They are defined as a set of special field lines that
  satisfy the condition of $\mathbf{\Omega \cdot B}=0$ at the light
  cylinder.} divide a polar cap into two distinct parts: the annular
gap region and the core gap region \citep[See Figure 1
  of][]{crab}. The annular gap is constrained between the critical and
last open field lines, and the core gap is around the magnetic axis and
within critical field lines \citep{AG10}.
The size of the polar cap decreases with increasing spin period,
  but can be quite large for MSPs and young pulsars.  The annular gap
width is correspondingly larger for short spin-period pulsars, and it
varies with the magnetic azimuth $\psi$.  For an anti-parallel
rotator and $\psi=0^\circ$, the radii of the annular polar region is
$r_{\rm ann}=r_{\rm pc}-r_{\rm core} = 0.26 R(\Omega R/c)^{1/2}$
\citep{AG10}, here $R$ is a pulsar's radius, and $\Omega$ is its
angular spin frequency.

Combining advantages of the outer gap and TPC models,
\cite{qiao04,qiao07} originally suggested the annular gap model, 
  which has been further developed by \cite{AG10,vela, crab}. The
site for generation of high energy photons is mainly located in the
vicinity of the null charge surface\footnote{It is defined as a
  surface where the Goldreich-Julian (GJ) charge density $\rho_{\rm
    GJ}$ \citep{GJ69} is zero. }. Being consistent with the physically
calculated spectra \citep{vela}, the Gaussian emissivities are
numerically assumed when simulating light curves. The key emission
geometry parameters $\alpha$ and $\zeta$ are not convincingly
confirmed so far, we adopt these two values from related
  literature if they exist. If there are not any reliable values, we
just use some random values according to theories of MSPs' magnetic
field evolution.
By hypothesizing reasonable emissivities and the magnetic inclination
angle, this model can make wide fan-like emission beams, thus well
reproduce the relevant light curves cut by a suitable viewing angle.

\subsection{Acceleration Electric Field}

\begin{figure}
\centering
\includegraphics[angle=-90,scale=.6]{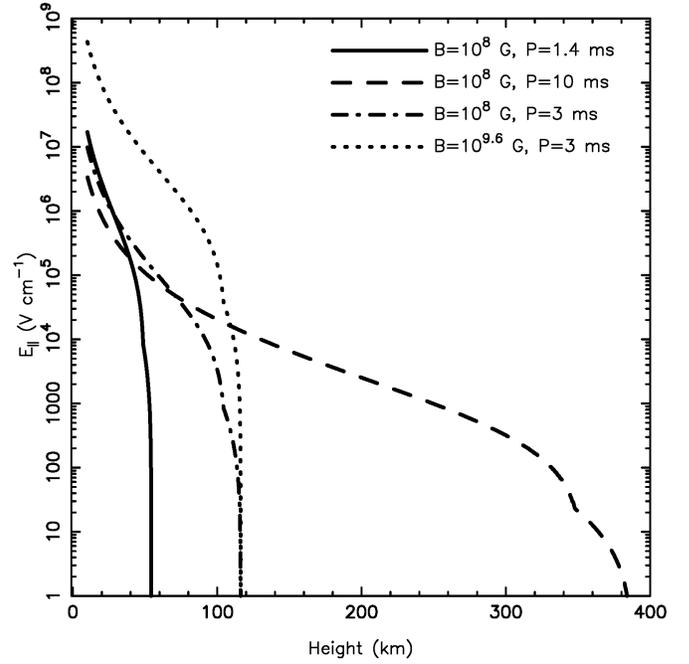}
\caption{The calculated acceleration electric field on an open field
  line in the annular gap are plotted for typical MSPs with four
  groups of parameters of surface magnetic field $B$ and spin period
  $P$. The electric field can extend from the pulsar surface to the
  null charge surface or even beyond it. The maximum electric field in
  the annular gap is sufficiently high to accelerate primary particles
  to ultra-relativistic energies.} \label{ELEC}
\end{figure}

Co-rotating charge-separated plasma is filled in a pulsar
magnetosphere \citep{GJ69,RS75}. When reaching some regions near the
light cylinder, charged particles can not exceed the speed of light
($c$), thus escape from the magnetosphere. To compensate the escaping
particles, the pulsar has to supply sufficient charged particles to
its magnetosphere. This dynamic process continuously happens, thus a
huge acceleration electric field is generated in the magnetosphere.
This is the general mechanism for acceleration electric field
($E_{\parallel}$), which is suitable for both young pulsars
\citep{vela,crab} and MSPs.
The charged particles with opposite signs are simultaneously exporting
from the annular gap and core gap, and satisfies the condition of
circuit closure in the whole magnetosphere.
%

We assume the flowing-out particles' charge density equals to the
local Goldreich-Julian density \citep{GJ69} at a radial distance of $r
\sim R_{\rm LC}$, here $R_{\rm LC}=\frac{cP}{2\pi}$ is a radius
  of light cylinder. The detailed calculation method and formulae are
presented in \cite{vela}. The acceleration electric fields of typical
MSPs are shown in Figure~\ref{ELEC} using four sets of surface
magnetic field $B$ and spin period $P$.
For MSPs with large or small values of $P$ and $B$, we found
  that, in all cases, the electric field in the inner region of the
  annular gap is sufficiently high ($E_\parallel \gtrsim 10^6\,{\rm
  V\, cm^{-1}}$).


The charged particles accelerated in the annular gap or core gap are
flowing out along a field line in a quasi-steady state. Using the
derived acceleration electric field, we can obtain the Lorentz factor
$\Gamma_{\rm p}$ of primary particles from the balance of acceleration
and curvature radiation reaction
\begin{equation}
\Gamma_{\rm p} = (\frac{3\rho^2 E_{\parallel}}{2e})^{\frac{1}{4}} = 2.36\times
  10^7{\rho_7}^{0.5} E_{\parallel,\, 6}^{0.25}, \nonumber
\label{gam_p}
\end{equation}
where $e$ is the charge of an electron, $\rho_7$ the curvature radius
in units of $10^7$\,cm and $E_{\parallel,\, 6}$ the acceleration
electric field in units of $10^6\, \rm V\, cm^{-1}$.

The primary particles are accelerated to ultra-relativistic energy
with typical Lorentz factors of $\Gamma_{\rm p} \sim 10^6 - 10^7$,
because of the huge acceleration electric field in the annular
gap. Since a lot of $\gamma$-ray photons are generated by the primary
particles via curvature radiation and inverse Compton scattering
processes, abundant ${\rm e}^\pm$ pairs are subsequently created
through two-photon annihilation and photon-magnetic-absorption
($\gamma$-B) processes.

\begin{figure*}[!htb]
\centering
\includegraphics[angle= 0,scale=.7]{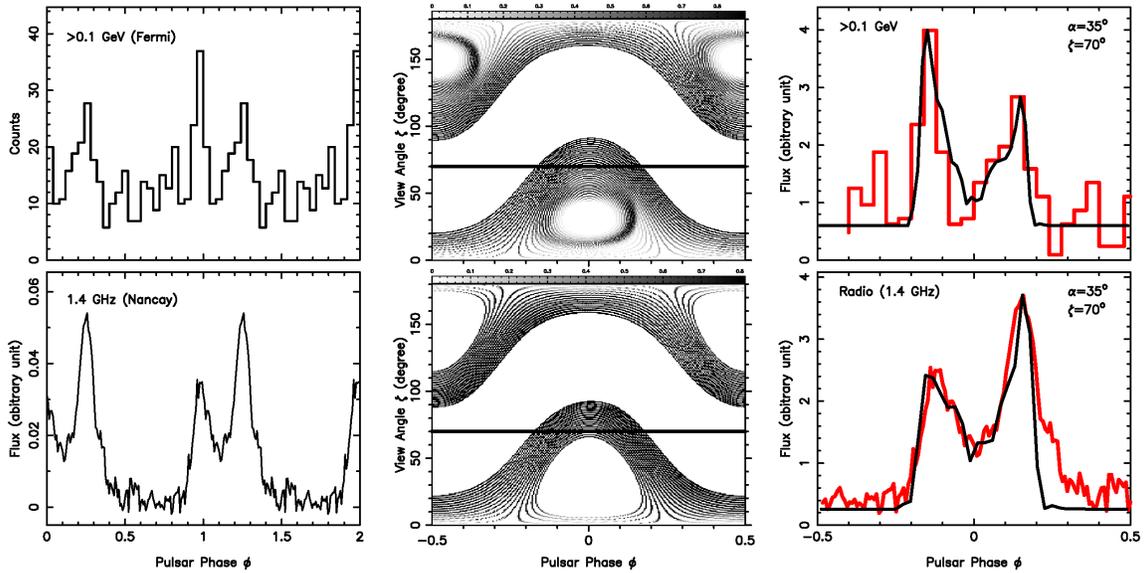}
\caption{Simulated radio and $\gamma$-ray light curves for PSR
  J0034-0534. The observations are presented in the left panels, and
  the radio data is taken from the {\it Fermi} ephemeris website
  http://fermi.gsfc.nasa.gov/ssc/data/access/lat/ephems/. The photon
  sky-map (middle panels) for an inclination angle of
  $\alpha=35^\circ$ and the corresponding simulated light curves (thin
  black lines in the right panels) for a viewing angle of
  $\zeta=70^\circ$ are also presented, using the single-pole annular
  gap model that well reproduce the observed phase-aligned light
  curves (thick red lines in the right panels). } \label{0034}
\end{figure*}

\section{Simulating Radio and $\gamma$-ray Light Curves for MSPs}

Thanks to the {\it Fermi}-LAT, we now know that some MSPs are
muti-wavelength emitters which have detectable radio and $\gamma$-ray
pulsed emission. According to the observations of phase lag ($\Delta$)
between radio peak and $\gamma$-ray peak \citep{catalog,kerr12}, MSPs
can be divided into four classes. PSR J0034-0534 represents a class of
MSPs which has phase-aligned light curves ($\Delta \sim 0$); PSR
J0101-6422 stands for another class which has moderate radio phase lag
$\Delta \sim 0.2 - 0.3$ with quite complex radio or $\gamma$-ray light
curves; PSR J0437-4715 stands for a third class which has larger radio
lag $\Delta \sim 0.43$ and PSR J1744-1134 is a fourth class of MSPs
whose $\gamma$-ray peak precedes the radio peak \citep{catalog}, we
will model this MSP in future when high signal-to-noise $\gamma$-ray
light curves are available.
We process {\it Fermi} Pass 7 data to derive the observed light
curves for three MSPs according to the radio timing solutions of MSPs
from Fermi Science Support Center
(FSSC)\footnote{http://fermi.gsfc.nasa.gov/ssc/data/access/lat/ephems/}.
We select events with energies of $> 0.1$\,GeV within 2$^\circ$ of
each MSP's position and with zenith angles smaller than
105$^\circ$. The key filter conditions for good time interval are rock
angle $<52^\circ$ and angsep(RA$_{\rm MSP}$, DEC$_{\rm MSP}$, RA$_{\rm
  SUN}$, DEC$_{\rm SUN}$)\,$ < 5^\circ$, where RA and DEC are right
ascension and declination respectively. Then we use tempo2
\citep{hobbs06} with {\it Fermi} plug-in to obtain the spin phase for
each photon. Finally we obtain the high signal-to-noise $\gamma$-ray
light curves for the three MSPs (see red lines in Figure 2, 3, 4).

A convincing model should have simple clear emission geometric picture
with reasonable input parameters, which can not only reproduce
multi-wavelength light curves for young pulsars but also for MSPs. In
this paper, we will jointly simulate radio and $\gamma$-ray light
curves for PSR J0034-0534, PSR J0101-6422 and PSR J0437-4715.
We briefly introduce the simulation method here. As shown in
  table 1, $\alpha$ and $\zeta$ are ``first-rank'' parameters, which
  are primarily adopted from the literature if they exist. When there
  are not any convincing values, we tend to use reasonable values of
  $\alpha$ from relevant theory on magnetic field evolution of pulsars
  \citep{ruderman91} and $\zeta$ is adopted randomly according to the
  simulated emission pattern (photon sky-map).
When $\alpha$ and $\zeta$ are fixed, several other model parameters
are carefully adjusted for the emission regions until the
  observed light curve of the corresponding band can be reproduced.
The model parameters for three MSPs are listed in table 1. 

\begin{table*}[htb]
\centering
\caption{Model parameters of multi-wavelength light curves for three
  MSPs
\label{tbl_1}}
%
\begin{tabular}{lccccccccc}
\hline \hline
Band & $\kappa$ & $ \lambda$ & $\epsilon$ & $\sigma_{\rm A}$ &
$\sigma_{\rm \theta,\,A}$ & $\sigma_{\rm C}$ & $\sigma_{\rm
  \theta,\,C1}$ & $\sigma_{\rm \theta,\,C2}$ \\
\hline
\multicolumn{9}{c}{J0034-0534 \;\; ($\alpha=35^{\circ}$;\,\,$\zeta=70^{\circ}$)} \\
\hline
$>0.1$\,GeV & 0.75  & 0.85 & 0.8 & 0.32 & 0.007 & 0.15  & 0.006  & 0.0052 \\
Radio & --  & -- & 0.58 & -- & 0.013 & -- & 0.00065  & 0.00053 \\
\hline
\multicolumn{9}{c}{J0101-6422 \;\; ($\alpha=30^{\circ}$;\,\,$\zeta=48^{\circ}$)} \\
\hline
$>0.1$\,GeV  & 0.50  & 0.85 & 0.8 & 0.3 & 0.009 & 0.15  & 0.006  & 0.006 \\
Radio & --  & -- & 1.5 & -- & 0.006 & --  & 0.0075  & 0.0078 \\
\hline
\multicolumn{9}{c}{J0437-4715 \;\; ($\alpha=25^{\circ}$;\,\,$\zeta=42^{\circ}$)} \\
\hline
$>0.1$\,GeV  & 0.1 & 0.1 & 1.3 & 0.03 & 0.002 & 0.02  & 0.001  & 0.001 \\
Radio & --  & -- & 1.2 & --  & 0.0004  & -- & 0.0002 & 0.0002 \\
\hline
\end{tabular}
\begin{tablecomments}
{ $\alpha$ and $\zeta$ are magnetic inclination angles and viewing
  angles; $\kappa$ and $\lambda$ are two geometry parameters to
  determine the peak altitude in the annular gap; $\epsilon$ is a
  parameter for the peak altitude in the core gap; $\sigma_{\rm A}$
  and $\sigma_{\rm C}$ are length scales for the emission region on
  each open field line in the annular gap and the core gap in units of
  $R_{\rm LC}$, respectively; $\sigma_{\rm \theta,\,A}$ is the
  transverse bunch scale for field lines in the annular gap;
  $\sigma_{\rm \theta,\,C1}$ and $\sigma_{\rm \theta,\,C2}$ are the
  bunch scale for field lines of $-180^\circ<\psi_{\rm s}<90^\circ$
  and $90^\circ<\psi_{\rm s}<180^\circ$ in the core gap,
  respectively. The detailed description of these symbols can be found
  in \citep{vela}.}
\end{tablecomments}

\end{table*}

\subsection{PSR J0034-0534}

\begin{figure*}
\centering \includegraphics[angle=0,scale=.7]{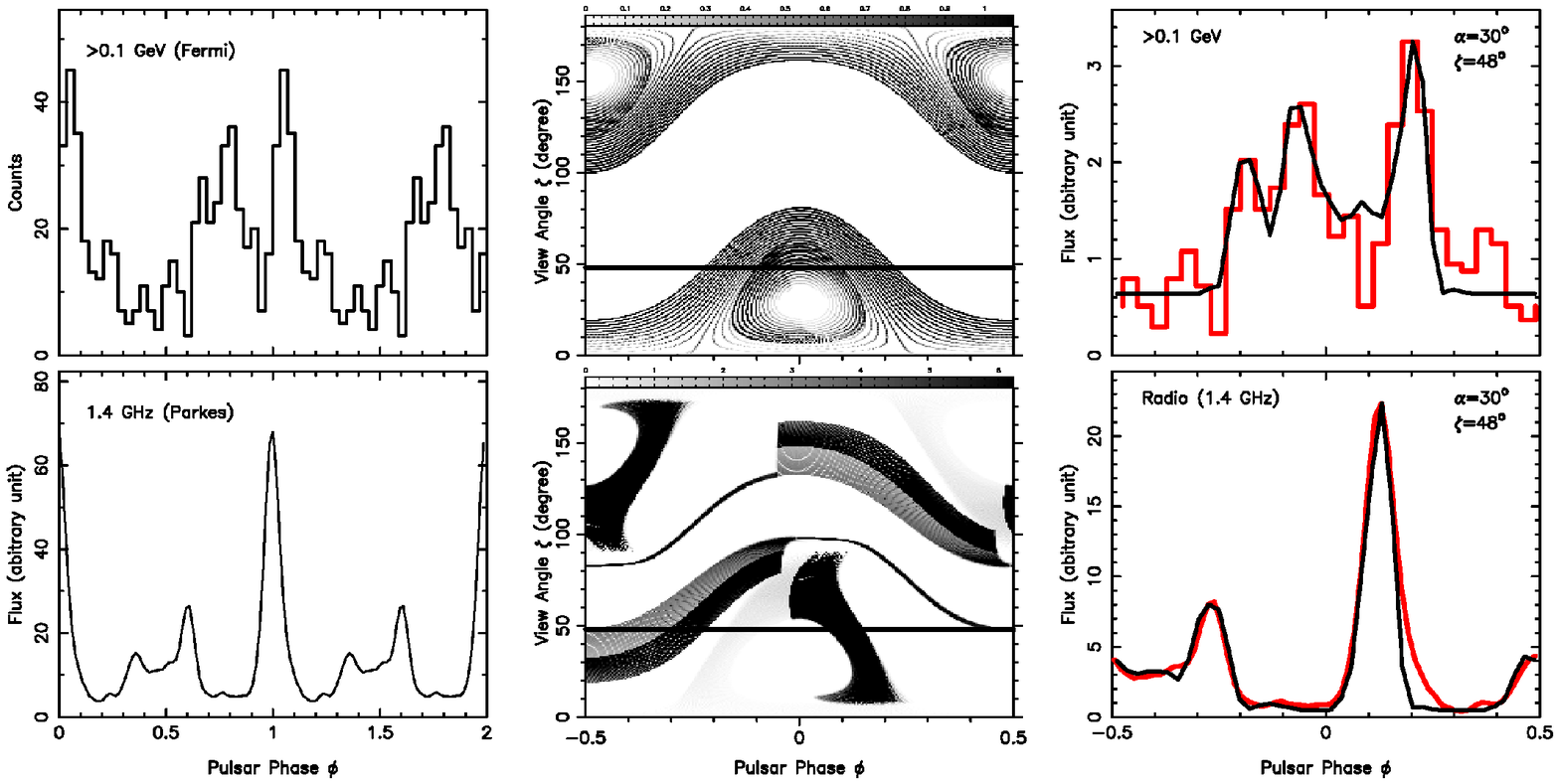}
\caption{Similar as Figure 2, but for PSR J0101-6422. The radio data
  is obtained from \citet{kerr12}. The inclination angle
  $\alpha=30^\circ$ and viewing angle $\zeta=48^\circ$ are used to
  model the complex radio and $\gamma$-ray light curves with moderate
  radio lag for this distinct MSP. }  \label{0101}
\end{figure*}

PSR J0034-0534 is the ninth $\gamma$-ray MSP detected by the {\it
  Fermi}-LAT \citep{abdo10}, and it has strong $\gamma$-ray and radio
pulsed emission with phase-aligned light curves. To reveal the
emission region of this MSP, we use the annular gap model to jointly
model the radio and $\gamma$-ray light curves. The simulation method
is the same as described in \S3.1 of \cite{vela}. The key idea
is to project all radiation intensities in both the annular gap
  and core gap to the ``non-rotating'' sky, considering some physical
effects (e.g. aberration effect and retardation effect). Here we use
numerical emissivities to speed up the calculations, since the
assumed emissivities are consistent with the physically calculated
spectra, as shown in Figure 8 of \cite{vela}.

From Figure 2, we find that both radio and $\gamma$-ray emission are
mainly generated in the annular gap region co-located at intermediate
altitudes $r \sim 0.24 - 0.56 R_{\rm LC}$, which leads to the
phase-aligned light curves. \cite{abdo10} used the TPC and outer gap
geometric models with $\alpha=30^\circ$ and $\zeta=70^\circ$ to obtain
the light curves for PSR J0034-0534, and they derived similar
conclusions that radio emission region extends from $0.6 R_{\rm LC}$
to $0.8 R_{\rm LC}$ and $\gamma$-ray region extends from $0.12 R_{\rm
  LC}$ to $0.9 R_{\rm LC}$. It is found that this MSP has a larger
transverse emission region for radio emission.
Moreover, \cite{venter12} developed the traditional outer gap and TPC
model, adopting the similar idea of numerically assumed emissivity of
piecewise-function, and derived the phase-aligned radio and
$\gamma$-ray light curves for three MSPs including PSR J0034-0534.
It seems that PSR J0034-0534 has off-peak pulsed $\gamma$-ray emission
up to 100\% duty cycle \citep{ackermann11}, which is not reproduced by
the annular gap model. This might be due to the lack of knowledge on
emission geometry and magnetic field configuration, and we will
further develop our model to study this MSP in detail in the
  future.


\begin{figure*}[!hbt]
\centering
\includegraphics[angle=0,scale=.7]{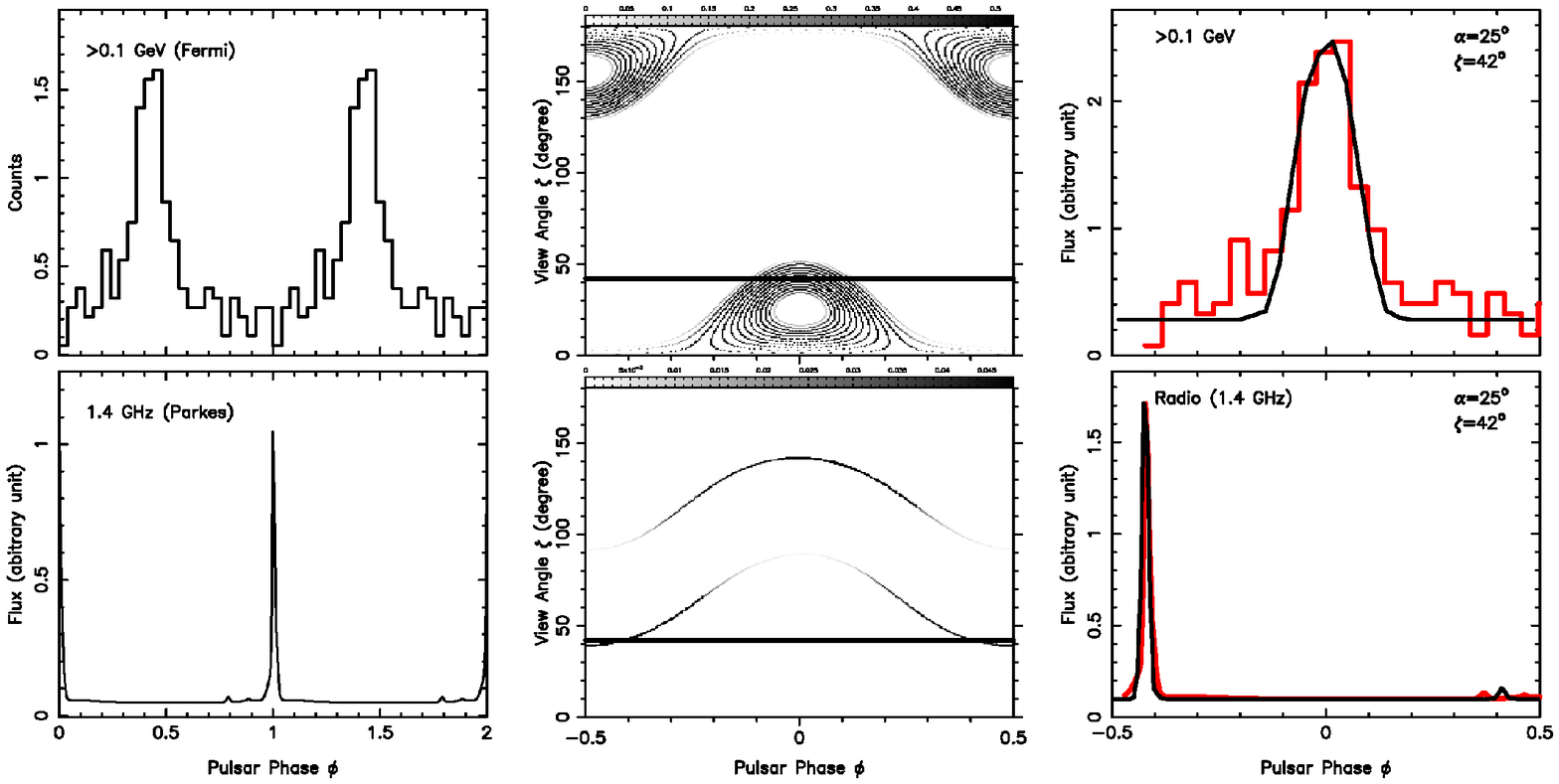}
\caption{Similar as Figure 2, but for PSR J0437-4715. From right
    bottom panel, the radio interpulse at phase $\sim -0.4$ and main
    peak at phase $\sim 0.4$ are reproduced by the annular gap model,
    and they are generated from a higher and narrower region in the
    annular gap region of the same magnetic pole as the $\gamma$-ray
    emission. The radio profile for this MSP is observed from Kunming
    40-meter radio telescope \citep{hao10}. The inclination angle
    $\alpha=25^\circ$ and viewing angle $\zeta=42^\circ$ are used.}
 \label{0437}
\end{figure*}

\subsection{PSR J0101-6422}

The observed light curves of PSR J0101-6422 have complex
  features: the $\gamma$-ray profile is likely to have three peaks;
while the radio profile contains three peaks that occupy nearly
  the whole rotation phase (see left panels of Figure 3). To well
model both light curves, we consider both cases of single-pole
and two-pole emission-picture with many attempts on a large number of
parameter space. The light curves of this MSP favor a single-pole
  emission picture with $\alpha=30^\circ$ and $\zeta=48^\circ$, and
the results are shown in Figure 3. For a viewing angle of
$\zeta=48^\circ$, the radio interpeak originates from the core
gap with high altitudes of $0.3 R_{\rm LC}$ to $0.7 R_{\rm LC}$; while
the other two peaks with a bridge (at the phases of $\sim -0.3$ and
$\sim 0.45$) originate from the annular gap with altitudes of
$0.45 R_{\rm LC}$ to $0.78 R_{\rm LC}$.
The $\gamma$-ray profile is similar, the interpeak (at phase of $\sim
0$) mainly originates from the core gap with high altitudes of
$0.32 R_{\rm LC}$ to $0.6 R_{\rm LC}$, while the other two peaks 
  originate from the annular gap with altitudes of $0.08 R_{\rm LC}$
to $0.4 R_{\rm LC}$.

According to the annular gap model, we note that the radio emission
from PSR J0101-6422 is quite asymmetric in magnetic 
  azimuth. This is possibly due to the special physical coherence
condition and propagation effects in the pulsar magnetosphere. We add
some discussions on radio emission in \S\,4.

\subsection{PSR J0437-4715}

PSR J0437-4715 is a very close MSP with a distance of 0.16\,kpc to the
Earth \citep{manchester05}, and has multi-wavelength emission.
\cite{chen98} suggested that PSR J0437-4715 was an aligned rotator. We
therefore adopted a relatively small magnetic inclination angle as
$\alpha=25^\circ$ and a reasonable viewing angle as $\zeta=42^\circ$
from the high-precision radio timing observations
\citep{van01,hotan06}. \cite{bogdanov07} also used the value of
$\zeta=42^\circ$ to successfully model the thermal X-ray pulsations of
this MSP. We apply our annular gap model to jointly simulate its radio
and $\gamma$-ray light curves, and the results are shown in Figure
4. We emphasize that the radio interpulse can be reproduced by our
model, although it does not precisely match the observations. The
radio emission including the main peak and interpulse originate
from a much higher and narrower region in the annular gap region with
high altitudes of $0.48 R_{\rm LC}$ to $0.57 R_{\rm LC}$; while the
$\gamma$-ray emission is generated in the annular gap region with
lower altitudes of $0.064 R_{\rm LC}$ to $0.15 R_{\rm LC}$ located in
the same magnetic pole. This leads to a large radio lag of $\Delta
\sim 0.43$.

\section{Conclusions and Discussions}

Pulsed $\gamma$-ray emission from MSPs has been observed by the
sensitive {\it Fermi}-LAT. Particularly, the specific pattern of radio
and $\gamma$-ray emission from the PSR J0101-6422 challenges the
outer gap and TPC models. A convincing model should apply not only
to young pulsars but also to MSPs. In this paper, we used the
annular gap model to jointly model the radio and $\gamma$-ray light
curves for three representative MSPs PSR J0034-0534, PSR J0101-6422
and PSR J0437-4715 with distinct radio phase lags.
For PSR J0034-0534 with phase-aligned radio and $\gamma$-ray light
curves, both bands are mainly generated in the annular gap
region co-located at intermediate altitudes.
For PSR J0101-6422 with complex radio and $\gamma$-ray pulse profiles,
we presented the best simulated results for this type of MSPs. The
radio interpulse originate from the core gap at higher
altitudes; while the other two radio peaks with a bridge 
  originate from the annular gap region. The interpeak 
  originates from the core gap region and the other two peak
  from the annular gap region.
For PSR J0437-4715 with a large radio lag, the radio emission
(including the interpulse) originates from a much higher and
narrower region in the annular gap region, and the $\gamma$-ray
emission has lower altitudes.


From simulations of these MSPs, the annular gap model favors a
single-pole emission pattern with small inclination angles ($\alpha
\lesssim 35^{\rm \circ}$) for MSPs.  This result is compatible with
theories of magnetic field evolution of MSPs in binaries: some
recycled pulsars tend to have aligning magnetic filed moment,
i.e. small magnetic inclination angle $\alpha$
\citep{ruderman91,chen98}. \cite{lamb09} presented a concrete
discussion on the $\alpha$ evolution of MSPs while they were
recycling in low mass X-ray binaries, and they note that the strong
interactions between spinning superfluid neutrons and magnetized
superconducting protons in a pulsar's core force the spin axis to
change. A MSP would be an aligned rotator ($\alpha \sim
0^\circ$) if the star's north and south magnetic poles are
forced toward opposite spin poles by the accretion disk, or would be
an orthogonal rotator ($\alpha \sim 90^\circ$) if both of
  the star's magnetic poles are forced toward the same spin pole.
Moreover, it is certainly unclear what happens when a MSP's
recycling process finishes,
\cite{young10} analyzed the new pulse width data of normal pulsars,
and found that the spin and magnetic axes would align when they spin
down due to dipole radiation and particle outflowing.

Several MSPs are simply assumed to be nearly orthogonal rotators
because they have a radio interpulse separated by a large phase of
$\gtrsim 180^\circ$ from its main pulse \citep{chen98}.
\cite{guillemot2} studied multi-wavelength light curves for two
  MSPs (PSR B1937+21 and PSR B1957+20), and found that fitting the
  radio polarization data of PSR B1937+21 favors an orthogonal
  rotator. As with most RVM fits the confidence area is large but the
  orthogonal configuration is further supported by the
  altitude-limited TPC and outer gap models.  
However, this is not universally true, at least in the case of the
annular gap model. The radio light curve of MSPs (e.g. PSR J0437-4715)
with an interpulse can be well explained by the annular gap model with
a small magnetic inclination angle.

By simulating light curves for MSPs in the annular gap model, we found
that the radio emission mainly originates from the high-altitude
narrow region in the annular gap region. The radio emission pattern
(photon sky-map) is patch-like in our model. The radio emissivities on
each field line (in the annular gap or core gap) vary slightly (nearly
uniform), but the case for $\gamma$-ray light-curve simulation is
quite different, they vary a lot. High energy ($\gamma$-ray and X-ray)
emission is generated by non-coherent radiation from relativistic
primary particles and pairs, while radio emission is suggested to be
generated by coherent radiation due to two-stream instability of
outward and inward pairs \citep{RS75}.
Here we focus on studying radio and $\gamma$-ray light curves for
MSPs, the concrete emission mechanism including polarization, spectral
properties and long-term stabilities of radio lags are however needed
to further study, considering the coherent condition and propagation
effects. \cite{han98} systematically studied the radio circular
polarization for pulsar integrated pulse profiles, and found that
sense reversals of circular polarization are observed across the conal
emission in some cases, unrestricted to core components. The
polarization property of high-altitude radio emission from both
annular and core gaps is a valuable subject to be investigated in
future.  We will keep on improving our model to present better
simulated light curves, especially for the phases of leading wing of
peak 1, trailing wing of peak 2 and off-peak pulses.


In this paper, we simulated radio and gamma-ray light curves for
3 MSPs. When high signal-to-noise data at other wavelengths is
  available in the future, we will re-simulate light curves and fit
the multi-wavelength phase-resolved spectra. In sum, the annular gap
model is a self-consistent model not only for young pulsars
\citep{vela,crab}, but also for MSPs, and multi-wavelength light
curves can be well explained by this model.

\acknowledgments
The authors are very grateful to the referee for the insightful and
constructive comments.
%
We would like to appreciate Matthew Kerr very much for giving us the
ephemeris and radio profile of PSR J0101-6422. YJD is supported by
China Postdoctoral Science Foundation (Grant No.: 2012M510047), and
partially supported by our institute project of ``Five Key Foster
Directions'' (Grant No.: Y22116EA2S). GJQ is supported by the
  National Basic Research Program of China (2012CB821800) and National
  Natural Science Foundation of China (10833003). DC is supported by
  National Natural Science Foundation of China (10803006, 11010250)
  and Advance Research Projects of Space Science (Grant No.:
  XDA04070000).

  {\sl Facilities: Fermi}

\end {document}